\documentclass[conference]{IEEEtran}
\IEEEoverridecommandlockouts
\usepackage{cite}
\usepackage{amsmath,amssymb,amsfonts}
\usepackage{algorithmic,algorithm}

\usepackage{graphicx}
\usepackage{parskip}
\usepackage{lipsum} 
\usepackage{hyperref}
\usepackage{textcomp}
\usepackage{xcolor}
\def\BibTeX{{\rm B\kern-.05em{\sc i\kern-.025em b}\kern-.08em
    T\kern-.1667em\lower.7ex\hbox{E}\kern-.125emX}}
\begin{document}

\title{Social Networks are Divulging Your Identity behind Crypto Addresses
\thanks{This report serves as the final documentation for \textit{Data Ethics and Data Security Seminar} in Winter Semester 2021 }
}

\author{\IEEEauthorblockN{ Shuo Chen}
\IEEEauthorblockA{\textit{Data Science Master Project} \\
\textit{ Ludwig-Maximilians-Universität München }\\
Munich, Germany \\
shuo.chen@campus.lmu.de}
\and
\IEEEauthorblockN{Shaikh Muhammad Uzair Norman}
\IEEEauthorblockA{\textit{Data Science Master Project} \\
\textit{Ludwig-Maximilians-Universität München}\\
Munich, Germany \\
u.noman@campus.lmu.de}
}

\maketitle
\thispagestyle{plain}
\pagestyle{plain}
\begin{abstract}
Cryptocurrencies, such as Bitcoin and Ethereum, are becoming increasingly prevalent mainly due to their anonymity, decentralization, transparency, and security. However, the completely public ledger makes the trace and analysis of each account possible as long as the identity behind the public address is revealed. Theoretically, social networks could make that happen when addresses are posted on social network platforms using accounts containing personal information. To verify such a possibility, we have collected public data from two major platforms, i.e. Twitter and Reddit, aiming to find potential privacy leakage behind the ETH public address. In the end, an easy-to-use retrieval application is also built for a better illustration. 
\end{abstract}


\section{Introduction}
Cryptocurrencies are taking the world by storm. In combination with other fanciful ideas such as Metaverse and NFTs, this is opening yet another way of perceiving wealth and assets. Such a craze is driven by several major appealing features of the blockchain technology they are based on, such as anonymity, decentralization, transparency, and security\cite{nofer2017blockchain}. 

\begin{figure}[ht!]
\centering
\includegraphics[width=0.4\textwidth]{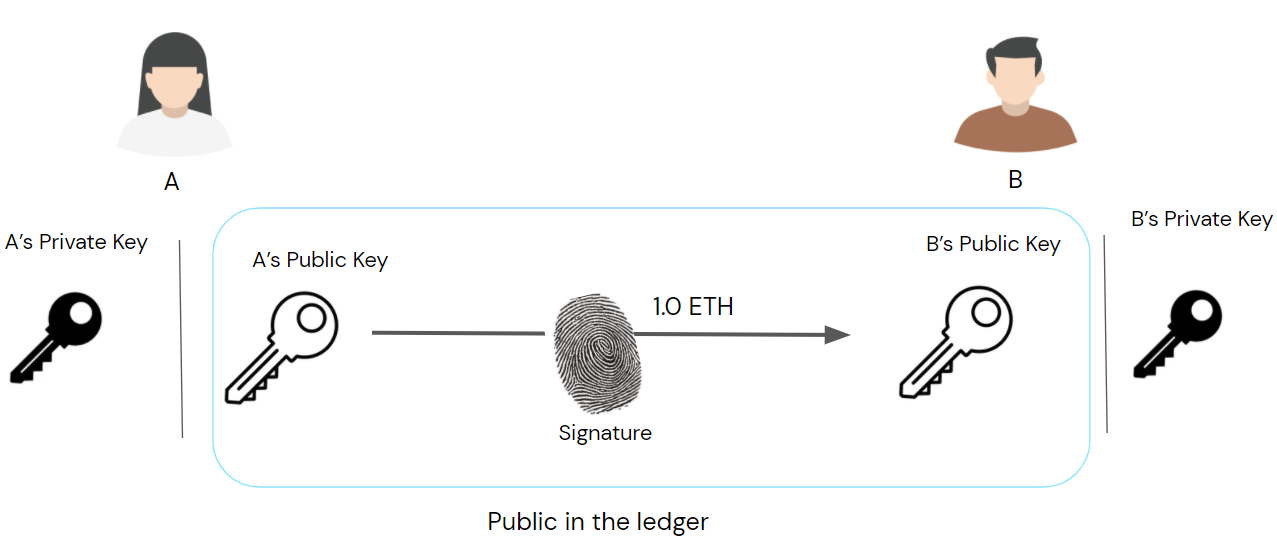}
\caption{A blockchain transaction illustration.}
\label{fig:process}
\end{figure}

The transaction mechanism on a blockchain is shown in Fig.\ref{fig:process}. For instance, if person A transfers 1 ETH to person B, then their public keys, the signature generated by the private key will be published in the ledger. Everyone has the ability to check this record. Given such a public ledger, due to the encryption algorithms, we cannot figure out their private keys but can verify the validity of this transfer. In this way, person A and B accomplish this transaction without knowing each other's identity.

However, transactions of cryptocurrencies are only considered pseudo-anonymous\cite{lansky2018possible}. Because every user who wants to make transactions needs a wallet address and anyone can view any transactions from the address due to decentralization. With everything being decentralized, it will be possible for everyone to access sensitive details as well, such as account balance and transaction history.  Starting from this idea, this report serves as a demonstration of how your addresses could reveal more personal information from many different sources.

Usage of such personal information is a double-edged sword. On one end, it risks data privacy, but on the other hand, it can help reveal fraudulent activities. For instance, companies like Chainalysis\footnote{\url{https://www.chainalysis.com/}}  have been identifying wallets that are linked to criminal activity, such as Darknet transactions. With their blockchain analysis software and other online, public clues, they were able to link transactions to real identities. Chainalysis’s most famous work was helping the FBI identify two rogue agents that had been stealing Bitcoins from the wallet of an online drug market operator\cite{bitcoinist}.

Cryptocurrencies are not fully regulated and therefore institution has to take a careful approach to let trustworthy and legit identities make use of their systems. With the increasing number of investors entering the scene, blockchain analysis can help banks and other financial institutions comply with KYC/AML – Know Your Customer and Anti-Money Laundering – laws.

Cryptocurrencies are pseudo-anonymous, therefore, researchers and developers are constantly working to make it more secure and anonymous. Zerocash \cite{sasson2014zerocash} and Zerocoin \cite{miers2013zerocoin} have been proposing a new digital currency with strong privacy guarantees. But currently, Bitcoin and Ethereum are still the two most popular cryptocurrencies \cite{marketsharing}, and again they are pseudo-anonymous. Besides, there is a more recent report in 2020 by Decrypt\footnote{\url{https://decrypt.co/}} \cite{decrypt} claiming that they analyzed 133,000 Ethereum names and their respective balances. Furthermore, they found that it is possible to identify several high-profile people, even if they were not using their real names. They were able to see business deals and watch people’s movements, just using the blockchain. Moreover, in \cite{bres2020blockchain}, reserachers profiled and deanonymized Ethereum users based on address activities using graph embedding algorithms. 

This work \cite{decrypt} motivates us and we think that besides the Ethereum names service, there is another more pervasive potential vulnerability threatening the anonymity, that is, social networks.

A famous quote is that "humans are the weakest link in cybersecurity" \cite{yan2018finding}. Cryptocurrency holders will inevitably use social networks for a myriad of scenarios where they need to post their public addresses. For example, they may leave their addresses 
to receive a donation or to participate in free-NFT activities. If not hiding properly, personal information is likely to be found through some matching. For instance, after posting your wallet address, even if the related account does not share any identifiable details, the user may have used the same username somewhere else. And there might be identifiable details in another post under that username. Also, the same email address used to sign up could be public and attackers could re-use the email to find out the account on another platform. All in all, it is not impossible to connect all the dots. 

To verify such a possibility, in our work, we leverage public information from social network platforms and cryptocurrency APIs to trace and unveil identities behind public wallet addresses. We also built a retrieval application\footnote{\url{http://45.76.88.238/}} based on the data we collected to illustrate our results. 

In the following content, detailed methods are discussed in Section \ref{Method}. Section \ref{results} demonstrates the results of our method and implementation. In the end, Section \ref{discussion}, \ref{future}, and \ref{conclusions} present our discussions, inadequacies, future work, and conclusions.  

%\lipsum[1-2]

\section{Methods}
\label{Method}
We aim to seek the real identity behind public cryptocurrency addresses through digging public data from social network platforms and blockchain APIs. More specifically, there are mainly two steps, \textit{data scraping and analysis}, and \textit{result display}. 
\subsection{Data Scraping and Analysis}
\label{analysis}
After a general review of the public APIs provided by several well-known social media platforms, we chose Reddit\footnote{\url{https://www.reddit.com/}} and Twitter\footnote{\url{https://twitter.com/}} as our main sources. Both platforms attract many crypto users and provide easy-to-use APIs with reasonable quotas. Besides, considering the popularity of Ethereum\footnote{\url{https://ethereum.org/en/}} and the convenience of its public APIs Etherscan\footnote{\url{https://etherscan.io/apis}}, we chose ETH as our main focus.

On Reddit, people share their addresses publicly for different purposes such as participating in promotions or bounties for free coins. We found that people frequently shared their ETH addresses on the fan pages of crypto pages which offered some promotional bonus to those sharing their addresses in the comments. Since these addresses are unique just like our email addresses, our idea is to search those addresses on Reddit to get more information on those particular users. By using several query keywords, we first looked for related posts, iterated through all the comments, and then used pattern matching to detect ETH addresses. As long as we found addresses in a comment or post, we recorded them with the user who post this information. Algorithm. \ref{alg:scrape-reddit} shows the details.

\begin{algorithm}
 \caption{Algorithm for data scraping on Reddit}
 \begin{algorithmic}[1]
 \renewcommand{\algorithmicrequire}{\textbf{Input:}}
 \renewcommand{\algorithmicensure}{\textbf{Output:}}
 \REQUIRE queries
 \ENSURE  reddit\_users, addresses
 \\ \textit{Initialisation} : reddit\_users, addresses
  \STATE load Reddit API key
  \FOR {query \textit{in} queries}
  \STATE get posts
  \FOR{post \textit{in} posts}
  \FOR{comment \textit{in} post}
  \IF {comment contains address}
  \STATE add user of the comment to reddit\_users
  \STATE add address to addresses
  \ENDIF
  \ENDFOR
  \ENDFOR
  \ENDFOR
 \RETURN  reddit\_users, addresses 
 \end{algorithmic} 
 \label{alg:scrape-reddit}
 \end{algorithm}

The next step is to find matches from other social media platforms and in our case, we chose Twitter. APIs from Twitter make searching for a typical query possible. Hence, after collecting enough addresses from Reddit, we turn to Twitter to find matches of these addresses. We searched posts containing these addresses and scraped the related account information (see Algorithm. \ref{alg:scrape-twitter}). In the next step, these matched Twitter accounts will be under scrutiny to find their public personal information such as their outlinks to Github or personal website, locations, and email addresses. In our work, all of this information is mainly mined from the public descriptions from each Twitter account via natural language processing.

\begin{algorithm}
 \caption{Algorithm for finding matches on Twitter}
 \begin{algorithmic}[1]
 \renewcommand{\algorithmicrequire}{\textbf{Input:}}
 \renewcommand{\algorithmicensure}{\textbf{Output:}}
 \REQUIRE reddit\_users, addresses, twitter\_rate\_limit
  \ENSURE  twitter\_info
 \\ \textit{Initialisation} : twitter\_info
  \STATE load Twitter API key
  \FOR {address \textit{in} addresses}
  \IF{address shows in Twitter}
  \STATE get personal information from the user who post the address
  \STATE add personal information to twitter\_info
  \IF{twitter\_rate\_limit is reached}
  \STATE program sleeps for 16 minutes
  \ENDIF
  \ENDIF
  \ENDFOR
 \RETURN twitter\_info
 \end{algorithmic} 
 \label{alg:scrape-twitter}
 \end{algorithm}

% how to analyze the description 

While accessing these APIs, we considered the security of our API tokens as we managed the project via Github\footnote{\url{https://github.com/}}. We applied the symmetric encryption algorithm Fernet \cite{zadka2019cryptography} to encrypt the API secret tokens, logging user name and password. There is only encrypted information stored in the code and the execution of our code requires our independent secret key file.  In this way, we protect the security of our API to some extent while hosting this project on Github. 

Last but not least, to gain an insight into the ETH account of each address, we accessed Etherscan APIs to obtain the account balance and transaction history. This could help clean the dataset by removing dead addresses with 0 balance and no transaction history. 

\subsection{Display Results}
% explain that we want a better way to display results 

\begin{figure}[ht!]
\centering
\includegraphics[width=0.45\textwidth]{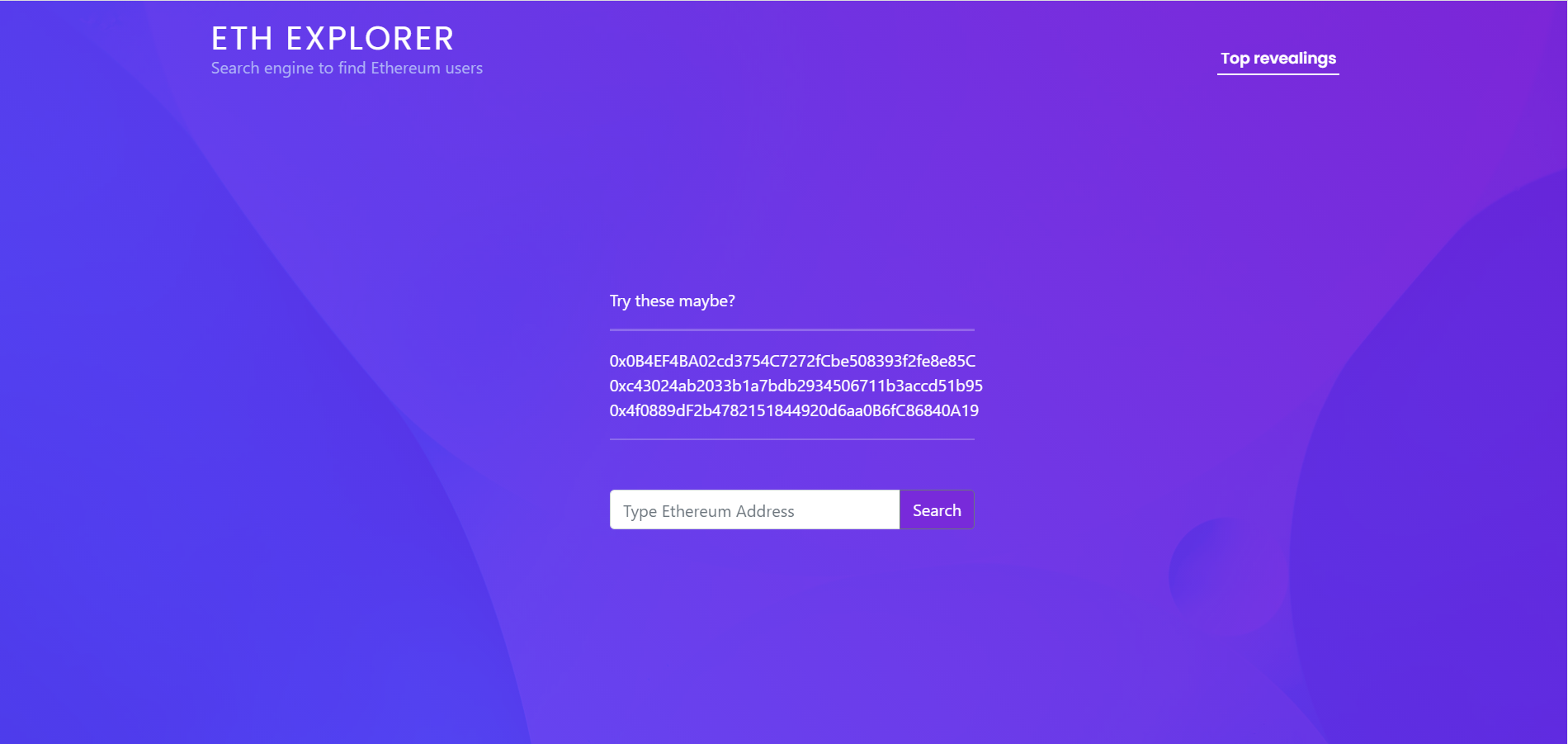}
\caption{Web application landing page.}
\label{fig:main}
\end{figure}
% a retrieval interface could also be further extended in the future with more functions 
We built a web application to query addresses, and obtain the details of the corresponding owner, such as the account balance and transaction history. The interface is made with Angular\footnote{\url{https://angular.io/}} and the data is from the result of Section \ref{analysis}.

We chose a tech-styled theme that goes well with the idea of cryptocurrency. The theme is using Bootstrap 5\footnote{\url{https://getbootstrap.com/}}. Bootstrap an is open-source front-end framework. It is famous for offering some very customizable options for designing websites. It also offers easy-to-use classes to make website responsive on different screens. So developers can control how the website will be shown on devices of different sizes which could be mobile, tablets, laptops, or larger screens. 

We also used the Underscore.js\footnote{\url{https://underscorejs.org/}}. Underscore.js is a javascript library that provides over 100 utility functions which include map, find, filter which can help with data of different data structures to be processed easily.

\begin{figure}[ht!]
\centering
\includegraphics[width=0.45\textwidth]{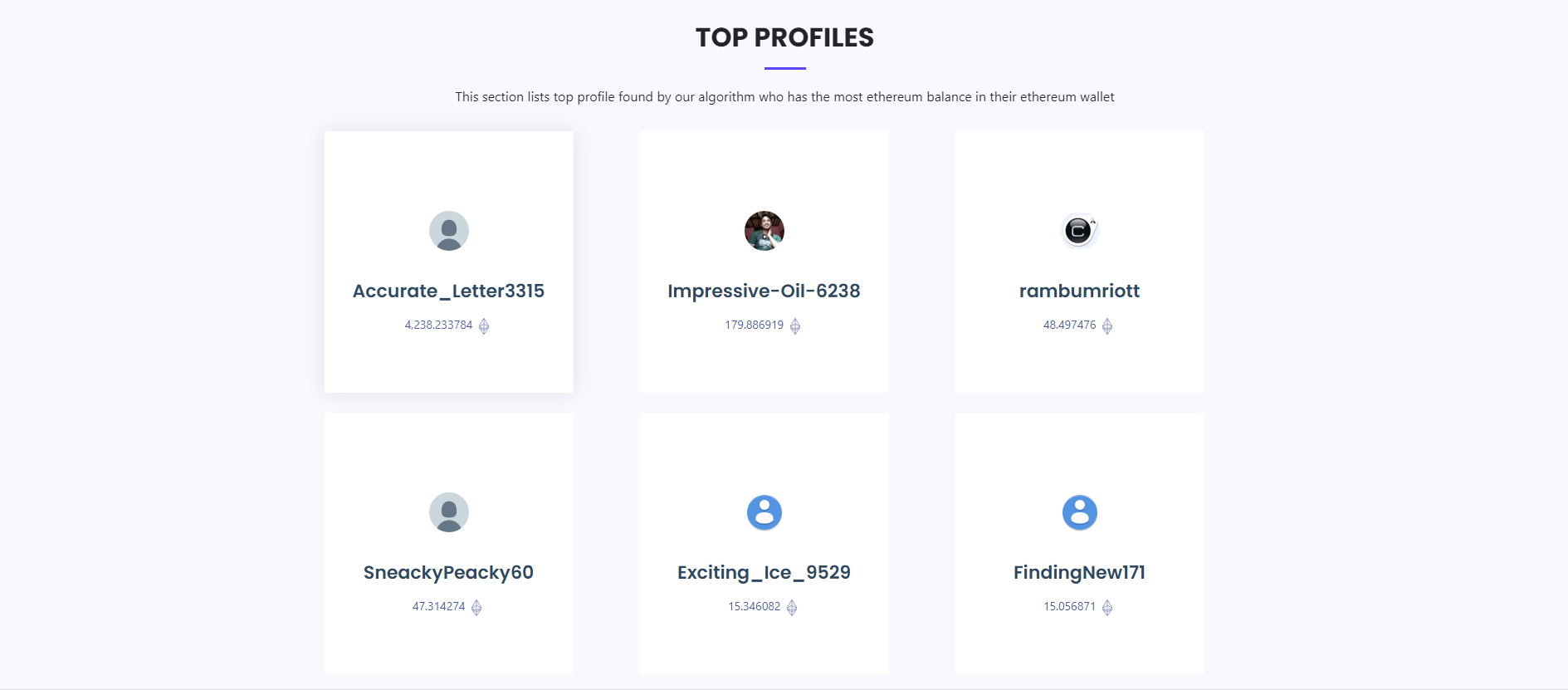}
\caption{Top profile section.}
\label{fig:top-prof}
\end{figure}

Web application boots up with the page as shown in Fig. \ref{fig:main}. With a very minimalistic header, a suggestive text to help users explore and play with our examples of addresses generated randomly, a most importantly a search bar to write Ethereum addresses. We came up with the idea of having a section to show the profiles of users having the highest Ethereum balance. The visual design of that section can be seen in the Fig. \ref{fig:top-prof}. 

\begin{figure}[ht!]
\centering
\includegraphics[width=0.5\textwidth]{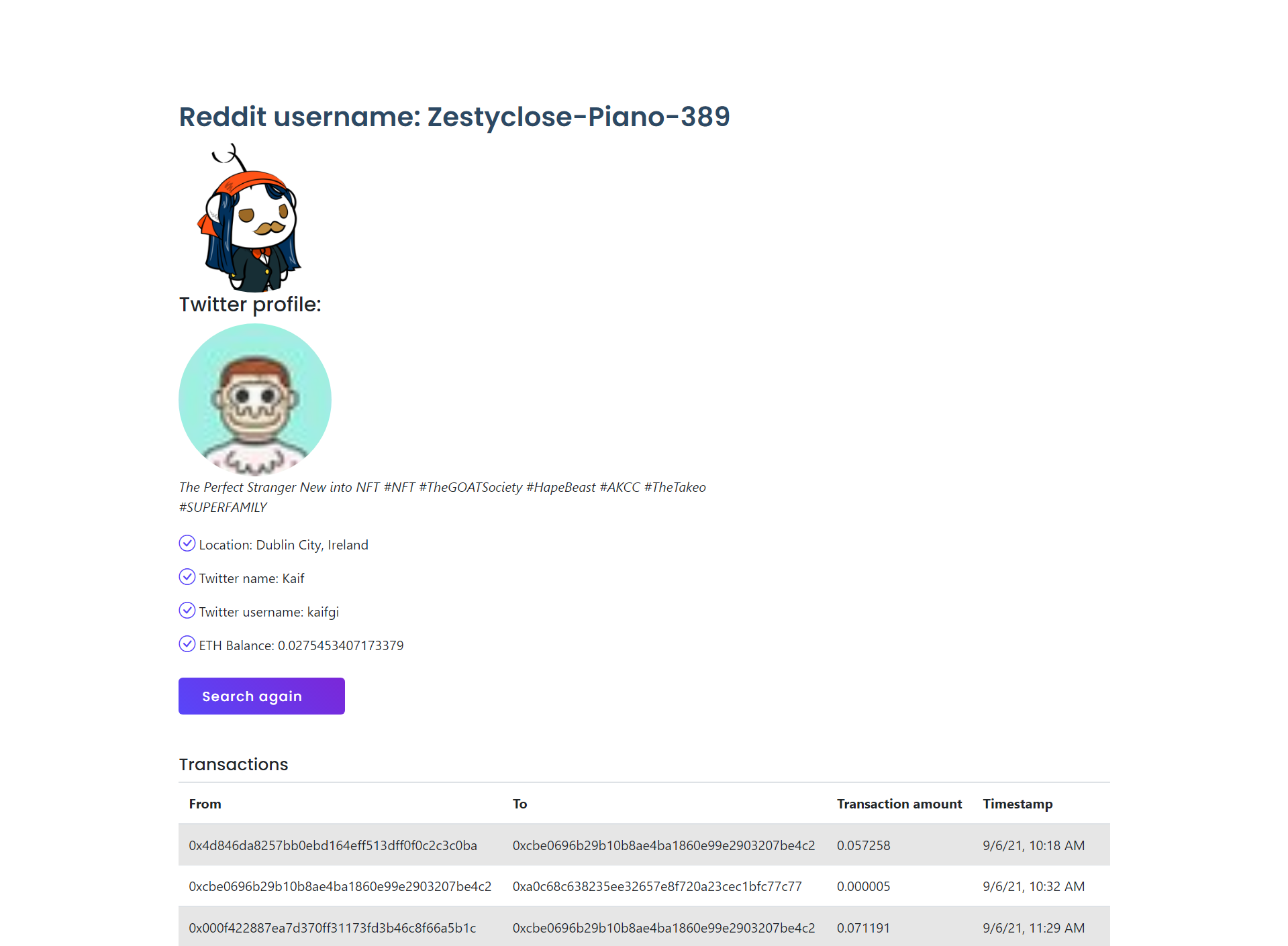}
\caption{Detailed profile page of a user.}
\label{fig:detail-pg}
\end{figure}

When the user types in the address in the search bar, or click on the suggested addresses, the app takes them to the detailed profile of that user showing their Reddit details and Twitter details (if exists in the database). Clicking the Twitter avatar also takes them to their existing twitter profile where you can have a closer look. We made use of Etherscan API to get transactional data attached to their addresses as well. So if transactional data of that user exists, we can also see a table below with all the transactions related to that user which contains to and from addresses, Ethereum amount, and the transaction timestamp as well. A profile having a complete detail found by our algorithm would look like Fig. \ref{fig:detail-pg}.

%\lipsum[1-8]
\section{Results}
\label{results}
%\lipsum[2-8]
\begin{table}[]
\caption{Number of records.}
\label{table:1}
\begin{tabular}{|c|c|c|c|c|}
\hline
Data Type & \begin{tabular}[c]{@{}c@{}}Scraping from \\ Reddit\end{tabular} & \begin{tabular}[c]{@{}c@{}}Dead \\ Addresses\end{tabular} & \begin{tabular}[c]{@{}c@{}}Active \\ Addresses\end{tabular} & \begin{tabular}[c]{@{}c@{}}Matches\\  of Twitter\end{tabular} \\ \hline
\begin{tabular}[c]{@{}c@{}}Nummber \\ of records\end{tabular} & 8551 & 5923 & 2634 & 1297 \\ \hline
\end{tabular}

\end{table}

Using the searching APIs from Reddit, we finally collected $8551$ different addresses and the corresponding users. Taking all these addresses and by comparing them with the query results from Etherscan APIs, we found that there are $2634 (30.8\%)$ active ETH account and 5923 $(69.2\%)$ dead ETH account (see Table. \ref{table:1}). After turning to Twitter for matches, we obtained $1297$ ETH addresses in total that have been found on Twitter . Within them, $571 (44\%)$ ETH accounts are active. 

As we seek personal information from the description of each Twitter account, we filtered matches without description and obtained $916$ records with descriptions. Among all the descriptions, we found $75$ with outlinks, $12$ with discord channel names and $10$ with emails (see Table \ref{table:2}). 

After taking a deeper look at these descriptions, we unveiled several users' identities. Considering privacy, we omitted several digits of sensitive information. A Reddit user with nickname \texttt{Houn**} has published his public address \texttt{0x8fa946f7c5d81fde14688a7a3eeb9b6cef7506**} for a free NFT distribution. Then we matched this address with a Twitter account with the same username \texttt{Houn**}. From his description, we get to know that he works in Canada, has a YouTube Channel with the name \texttt{All** BMWDurh**}, and owns an Instagram account with the name \texttt{all**bmwdurh**}. Under his Instagram introduction, we found that his real name is \texttt{Al*** L**}, and he works at \texttt{B** Can***}. 

A second example follows the same manner. Reddit user \texttt{Desperate-Spite-****} published his address \texttt{0xd6Db06b4FB93Be68F93Dc75176D4D2A538B294**} under a post, then we linked this address with a Twitter account \texttt{Olani** Muyi** Mirac**} and found that he is a photographer from Ibadan in Nigeria and has an Instagram account \texttt{@moremiracl**}. We also found his phone number which is \texttt{+23470608925**}.

The last example is of a programmer. \texttt{rambumrio**} post his address \texttt{0xbc7250c8c3eca1dfc1728620af835fca489bfd**} on Reddit. We found the match account \texttt{C0nw0**} on Twitter. Following the account description, we know his Github \texttt{C0nw0**} and that he is called \texttt{Con** Mcknig**} and works for a \texttt{Football Club}. He is also a developer mainly interested in front-end development. He seems rich in his ETH account and holds 48 ETH coins in total.

\begin{table}[]
\caption{Details of records with descriptions.}
\label{table:2}
\begin{tabular}{|c|c|c|c|c|c|}
\hline
Data Type & \begin{tabular}[c]{@{}c@{}}Matches\\  of Twitter\end{tabular} & \begin{tabular}[c]{@{}c@{}}with \\ Descriptions\end{tabular} & \begin{tabular}[c]{@{}c@{}}with\\  Links\end{tabular} & \begin{tabular}[c]{@{}c@{}}with \\ Emails\end{tabular} & \begin{tabular}[c]{@{}c@{}}with \\ Discord\end{tabular} \\ \hline
\begin{tabular}[c]{@{}c@{}}Nummber \\ of records\end{tabular} & 1297 & 916 & 75 & 10 & 12 \\ \hline
\end{tabular}
\end{table}
%\newpage
\section{Discussions}
\label{discussion}
%\lipsum[3-8]
% what we have done, a summary 
We utilized publicly available data via social network APIs to unveil personal-related information behind the public wallet addresses of Ethererum. The main goal of our projects has two folds. One is to alert all cryptocurrency users that humans are the weakest link in cybersecurity \cite{yan2018finding}. An unconscious post and remark on social media might annihilate all the security from intricate and convoluted decryption algorithms. The situation could get worse when it is related to money or asset-related matters. The second point is to demonstrate the possibility to trace and recognize malicious accounts used for illegal and anti-human-rights purposes.

We have constructed our dataset with more than 8 thousand records, in which some contains personal info from Twitter and Reddit and also ETH account information. Such dataset is built totally from public accessible APIs and could be easily extended in many ways, for example, by adding more search keywords on Reddit or in an extreme case, iterating all crypto related posts on Reddit and Twitter to dig out all the public addresses with their user information. Also, other popular social media platforms could be included such as Facebook\footnote{\url{https://www.facebook.com/}} , Instagram\footnote{\url{https://www.instagram.com/}} , or more personal-related website like 	LinkedIn\footnote{\url{https://www.linkedin.com/}}. By scraping public addresses and matching different users on various social network platforms, we cannot imagine how massive the final dataset could be, but we are sure there must be countless addresses which could reveal personal identity-related information.

Such a massive dataset visualized and served on a easy-to-use web interface could be a powerful tool from a big data and statistics related perspective too. For investigations of malicious anonymous transactions, investigators could first turn to such a tool to check whether there are readily available leaked personal information or related accounts. Normal cryptocurrency users could treat such a tool as a detection of their personal information leakage and can then be more careful.

With the advancement in Web 3.0 which is more about decentralization and having a “decentralized identity” of everyone. This tendency of sharing addresses and wallet details would be common in the future. One has to be really careful with the data they are sharing or agreeing to share data between applications or social media platforms. 

% approaches to prevent such attacks

Fortunately, there are some good projects being evolved that take care of financial privacy. Examples of such projects are ZCash\footnote{\url{https://z.cash/}} \cite{sasson2014zerocash} and Tornado Cash\footnote{\url{https://tornado.cash/}}. 
\\
\\
Tornado Cash makes private transactions which means that users can make ether transactions without disclosing the amount or leaving trails for potential privacy attacks. Tornado Cash and ZCash both use the concept of zero-knowledge proofs, which is a cryptography tool to sign a transaction without revealing personal details \cite{decrypt-privacy}. Tornado claims, “Whenever ETH is withdrawn by the new address, there is no way to link the withdrawal to the deposit, ensuring complete privacy”. While on the other hand, ZCash offers multiple transaction modes allowing people to selectively share address and transaction information for auditing or regulatory compliance.

% privacy of crypto compared to the traditional banking system
Cryptocurrencies are certainly a good step towards the more easy-to-use and less transaction costly option for trading and exchanging assets. Banking systems have been quite troublesome for many due to the paperwork, complex and hideous account policies. Cryptocurrencies have taken these manual human-dependent businesses to a more public and transparent environment. That obviously has its pros and cons. So on the one side, Cryptocurrencies offer faster and seamless transaction options without many hassles or involving bureaucracy, but on the other hand, it is complex for everyone to understand from the technical perspective and since decentralization is the core selling point, we cannot expect this feature to be removed from cryptocurrencies. But we can expect more development in the financial privacy sector and more options like ZCash or Tornado Cash to be available.

\section{Future Work}
\label{future}
% our limitation and future work 
We should point out that there are several shortcomings in our project. We summarize several points here and add them to the future agenda. 

Firstly, our dependence on public APIs provided by social network platforms hinders the flexibility of searching for more information. These APIs are either limited in speed and number of access per second or not comprehensive enough to provide all relevant information. While it is totally understandable considering the overhead costs and the content ownership, we will turn to more flexible web scraping technology such as Selenium \cite{selenium}\cite{chaulagain2017cloud} in the future.  

Another inadequacy is the lack of records and diversity of social platforms. We only scraped data from Reddit and Twitter and just focused on posts and comments. The result will be more promising if more platforms are taken into accounts such as Facebook, Instagram, and forums, especially those in Darknet. Also, deeper backtracking to the history of each account could also bring more relevant information regarding personal information.

Besides, a major part of our scraped addresses is used to receive or distribute NFTs rather than real transactions. This is mainly caused by the limited query words. Such a biased dataset could lead to noisy results as these accounts are not used for general purposes like purchasing but only collecting NFTs. 

In the future, we can also link Ethereum Name Service (ENS) with our current dataset to get more details related to those addresses. For example, we could know if someone's ENS domain is registered to that address or not, and retrieve other information such as his BTC, LTC, DOGE addresses, if available. Moreover, we can also get his email, GitHub, Discord, Telegram ids as well if he has made them public.

Last but not least, in the future, more complex analysis methods could also be implemented. For instance, intersection analysis of transactions from other addresses whose personal information is available could bring more insight into how these users connect with each other. We can also employ network science into these closely connected transactions to detect fraudulent individuals or groups.

\section{Conclusion}
\label{conclusions}
% our conclusions 
This report discusses that the tendency of posting addresses over social media is quite common in practice and how someone's address could reveal the personal information of individuals. This paper also serves as an awareness that each and every transaction could be peeked into by someone on the decentralized network which could be a potential security leak. We discussed our approach to get addresses and information from Reddit and combine it with other social media platforms such as Twitter.We were also been able to get their transaction history on that wallet address as well.

Through our website, we demonstrated the possibility of getting people having the most Ethereum amount in their wallets. Our website presents a simple search-engine use case of that data and provides a query interface through which one can input the addresses and get the details attached to that address.

\section*{Acknowledgment}
We would like to thank Jan Schmidt, Fabio Genz and Prof. Dieter Kranzmüller for hosting this seminar and also for constructive suggestions during preparation. All APIs used in this project are free of charge and publicly available, hence we are grateful to Reddit and Twitter for their openness. 

%\newpage

\bibliographystyle{IEEEtran}
\bibliography{references}  
\end{document}